# Gap Anisotropy, Spin Fluctuations and Normal-State Properties of the Electron Doped Superconductor $Sr_{0.9}La_{0.1}CuO_2$

G. V. M. Williams[1,2], R. Dupree[3], A. Howes[3], S. Krämer[1], H. J. Trodahl[4], C. U. Jung[5], Min-Seok Park[5], Sung-Ik Lee[5].

[1] 2. Physikalisches Institut, Universität Stuttgart, D-70550 Stuttgart, Germany

[2] Industrial Research Limited, P.O. Box 31310, Lower Hutt, New Zealand.

[3] Department of Physics, University of Warwick, Coventry, CV4 7AL.

[4] School of Chemical and Physical Sciences, Victoria University, Private Bag, Wellington New Zealand.

[5] National Creative Research Initiative Center for Superconductivity and Department of Physics, Pohang University of Science and Technology, Pohang 790-784, Republic of Korea

*20 December, 2001*

## ABSTRACT

We report the results from a thermopower and a Cu nuclear magnetic resonance (NMR) study of the infinite $CuO_2$ layer electron doped high temperature superconducting cuprate (HTSC), $Sr_{0.9}La_{0.1}CuO_2$. We find that the temperature dependence of the thermopower, S(T), is different from that observed in the hole doped HTSC. In particular, we find that dS(T)/dT is positive above ~120 K. However, we show that S(T) can still be described by the same model developed for the hole doped HTSC and hence S(T) is not anomalous and does not imply phonon mediated pairing as has previously been suggested. The Cu NMR data reveal a Knight shift and spin lattice relaxation rate *below* $T_c$ that are inconsistent with isotropic s-wave pairing. The Cu spin lattice relaxation rate in the normal state, however, is Curie-Weiss like and is comparable to that of the optimally and overdoped hole doped HTSC, $La_{2-x}Sr_xCuO_4$. The magnitude of the Knight shift indicates that the density of states at the Fermi level is anomalously small when compared with the hole doped HTSC with the same $T_c$, indicating that the size of $N(E_f)$ is of little importance in the HTSC. We find no evidence of the normal state pseudogap that is observed in the hole doped HTSC and which was recently reported to exist in the electron doped HTSC, $Nd_{1.85}Ce_{0.15}CuO_4$, from infrared reflectance measurements.

**Introduction**

One of the major difficulties in developing a consistent theoretical model of superconductivity in the high temperature superconducting cuprates (HTSC) is understanding the magnetic and electronic behavior of the hole doped and electron doped HTSC. It is now generally agreed that the hole doped HTSC (HDHTSC) have a d-wave superconducting order parameter [1,2], show a normal state pseudogap for low hole doping [3] and exhibit a non-Korringa relationship between the Cu spin-lattice relaxation rate and the Cu Knight shift [4]. This non-Korringa behavior has been attributed to the Cu spin-lattice relaxation rate being dominated by coupling to antiferromagnetic fluctuations [4]. The situation in the electron-doped HTSC (EDHTSC), based on $R_{2-x}Ce_xCuO_4$ (R is a rare earth), is not so clear. There are conflicting reports concerning the symmetry of the superconducting order parameter [5-10]. Furthermore, it has recently been reported that a pseudogap exists when superconductivity is suppressed by high magnetic fields but only for temperatures less than ~25 K [11-13] or there exists a pseudogap in the normal state for temperatures up to 300 K [14]. The understanding of the spin dynamics is hampered by the magnetic rare earth atom, which is believed to dominate the spin-lattice relaxation. It is known that the nuclear quadrupole frequency is anonymously small when compared to that of the HDHTSC, which is not understood [15].

Previous reports of NMR measurements on the infinite layer EDHTSC, $Sr_{1-x}La_xCuO_2$ [16,17,18], have shown that the Cu spin lattice relaxation rate in the normal state increases with decreasing temperature, although the absolute values of the Cu spin lattice relaxation rate are different in these studies. This compound has the advantage that it does not contain magnetic rare earth ions that can affect the NMR data. Furthermore, the superconducting transition temperature, $T_c$, is much higher (~40 K [19]) than that in the other EDHTSC, $R_{2-x}Ce_xCuO_4$, where superconductivity exists for lower temperatures of <25 K and the occurrence of superconductivity and the Meissner fraction depends upon aggressive annealing at high temperatures in a reducing atmosphere. The one disadvantage of $Sr_{0.9}La_{0.1}CuO_2$ is that it must be made using a high pressure synthesis technique and incorrect synthesis can lead to site disorder, the partial formation of the 1D antiferromagnetic insulator, $SrCuO_2$, as well as the antiferromagnetic insulator, $La_2CuO_4$.

Interest in the infinite layer compound has recently been revived by a report that the carrier concentration of $CaCuO_2$ can be altered by field effect doping resulting in hole doped or electron doped superconductors [20]. It is therefore important to compare the normal and superconducting state properties of the infinite layer superconductors with those from the HDHTSC. For example, tunneling spectra, and the rate at which $T_c$ is suppressed by Zn and Ni, have been interpreted in terms of a s-wave superconducting order parameter [21-24], which would appear to suggest a different superconducting pairing mechanism when compared with the HDHTSC. Furthermore, a thermopower study has concluded that the superconducting pairing mechanism is phonon mediated [25].

In this paper we report the results from a Cu NMR study and a variable temperature thermopower study of the infinite layer EDHTSC, $Sr_{0.9}La_{0.1}CuO_2$. Dense high quality samples of this compound have recently been synthesized with a high superconducting transition temperature of 43 K [23]. We compare our results with those from previous measurements on the HDHTSC.

**Experimental Details**

The sample for this study was prepared using the high-pressure technique described elsewhere [23,26]. The dc magnetization measurements were made using a SQUID magnetometer, where the remnant field was ±0.1 mT. The data were not corrected for demagnetizing effects. Variable temperature thermopower measurements were made using the standard temperature differential technique. X-ray diffraction measurements confirmed that the sample was single phase.

Cu NMR measurements were made on both unorientated powder and c-axis aligned powder samples in resin. Powder spectra were obtained at fields of 5.6 T, 8.45 T, and 14.1 T, point-by-point using the $\pi/2$-$\tau$-$\pi$-$\tau$ Hahn pulse sequence, where $\pi/2$ was between 2 μs and 10 μs and $\tau$ was between 30 μs and 40 μs. The second half of the echo was Fourier transformed and then integrated, or summed, to obtain the intensity at each frequency. As the probes used to obtain the spectra had a Cu background, data was not taken near this frequency. The temperature dependence of the shift and the $^{63}$Cu spin lattice relaxation time, $^{63}T_1$, was measured at 9 T in a probe with no Cu background. The NMR shift was referenced to an aqueous CuCl solution. The $^{63}$Cu spin-lattice relaxation rate, $T_1^{-1}$, was obtained using both the inversion recovery technique and from the magnetization recovery after the application a saturating comb of $\pi/2$ pulses. In both cases the magnetization recovery for the I=3/2 Cu nuclei for magnetic relaxation ($\Delta m=\pm 1$) is [27],

$$M(\tau) = M_0 \left[ 1 - \Gamma \left( A \exp\left(-\frac{\tau}{T_1}\right) + B \exp\left(-\frac{3\tau}{T_1}\right) + C \exp\left(-\frac{6\tau}{T_1}\right) \right) \right] \quad (1)$$

where $\tau$ denotes the time between the inversion pulse or the last saturation pulse, $\Gamma$ is 2 for perfect inversion and 1 for the application of a perfect $\pi/2$ saturating comb. The coefficients $A$, $B$ and $C$ depend on the initial excitation conditions. For the inversion or saturation of the central ±1/2 transition without exchange between the other levels, it can be shown that $A = 0.1$, $B = 0$ and $C = 0.9$. In the presence of fast exchange processes these coefficients change to $A = 0.4$, $B = 0$ and $C = 0.6$. It was found that the second scenario better fitted the data. The excitation of the +1/2↔+3/2 and -1/2↔-3/2 transitions result in different $A$ and $C$ values and a nonzero $B$ value.

**Results and Analysis**

The temperature dependent thermoelectric power, S(T), shown in figure 1 is zero in the superconducting state, as required for a strictly perfect superconductor. It becomes finite and negative at the transition temperature of 44 K determined from susceptibility measurements (see figure 1 inset). The thermopower continues to become more negative until reaching a maximum negative value near 110 K and then falls toward zero.

It is interesting to compare this behavior to that in the HDHTSC, in which a very general trend has emerged. In those materials the thermopower just above the superconducting transition temperature falls with a nearly constant (and approximately universal) slope [28,29]. The magnitude near the superconducting transition temperature depends on the doping level, which has led to its use as a gauge of the hole concentration [30]. The only superconducting cuprates to deviate from this general behaviour are those with conducting chains and/or ribbons, which are thought to make their own positive slope contribution to the thermopower [31]. An interpretation of this generic behaviour is available based on the doping dependence of the Fermi surface, and a conventional treatment of the charge carrying quasiparticles [32]. Within this picture the thermopower is comprised of a diffusion contribution, linear in temperature with a sign and magnitude that is determined by the Mott formula, and a phonon drag contribution, $S_g$, which is given by,

$$S_g = A(C_a N |e|)(W_{pe}/W_p). \quad (2)$$

Here $C_a$ is the volumetric heat capacity of the acoustic phonons, $N$ is the concentration of carriers, $e$ is the electronic charge, $W_p$ is the scattering rate for phonons and $W_{pe}$ is the contribution from events in which the phonons are scattered by electrons. The sign and magnitude of the constant $A$ is determined by a balance of scattering events in which the electron traverses filled vs. empty states in reciprocal space.

The generic behavior of the thermopowers in the HDHTSC then fits within this picture as being a contribution of a diffusion term and the temperature independent phonon drag contribution imposed by the heat capacity at temperatures above the acoustic phonon maximum. This interpretation implies that there is no temperature dependence in the ratio $W_{pe}/W_p$, consistent with a picture in which the temperature dependent anharmonic phonon-phonon scattering does not significantly limit the phonon lifetime. Similar behavior is occasionally found also in conventional metals [33], although it is more common to find that the phonon drag contribution falls as 1/T above the Debye temperature, where anharmonic scattering limits the phonon lifetime [34,35]. Thus, in conventional metals one usually finds a low temperature thermopower showing a linear temperature diffusion component and a cubic temperature phonon drag component, while at higher temperatures the drag contribution is inversely dependent on the temperature. It can be seen in figure 1 that the thermopower in the present case conforms to the more common behavior,

$$S(T) = a_1 T + a_2 T^3, \qquad (3a)$$

for low temperatures, and,

$$S(T) = a_1 T + a_3/T, \qquad (3b)$$

for high temperatures. The linear diffusion term coefficient, $a_1$, and the low and high temperature phonon drag coefficients, $a_2$ and $a_3$, are fitted as $a_1 = -0.00122 \mu V K^{-2}$, $a_2 = -430 \mu V K^{-4}$ and $a_3 = -1.15 \times 10^{-5} \mu V$. The magnitudes found for the various contributions are entirely reasonable when compared to either theoretical expectations or to data on the HDHTSC or conventional metals [34]. The high temperature 1/T attenuation of $S_g$ caused by the ratio $W_{pe}/W_p$ is however weak, suggesting that even here the phonon lifetime is limited more strongly by electrons than by anharmonicity. There is no evidence of any deviation from the expected behavior that might suggest the development of a pseudogap.

Finally, we note that there is an earlier report of thermopower measurements on thin films of the EDHTSC, $Sr_{0.9}Nd_{0.1}CuO_2$, where $T_c$ varied from 24 K to non-superconducting [25]. In that report the thermopower data were also fitted to a sum of linear temperature and inverse temperature terms, though the temperature range chosen (<90 K) was inappropriate for assigning the inverse-temperature behavior to a phonon drag contribution. Consequently, the conclusion that the low temperature thermopower data (<90 K) from the thin film samples implies phonon mediated pairing in the EDHTSC is also inappropriate.

We now consider the Cu NMR data. We show in figure 2 that the room temperature $^{63}$Cu NMR powder spectra are narrow and nearly independent of applied magnetic field when scaled by the Larmor frequency. The resonance from the $^{65}$Cu isotope occurs at a higher frequency and is not shown. The $^{63}$Cu NMR spectrum at 14.1 T contains a narrow region of ~1.4 MHz in extent and a broad weaker region extending to ±6 MHz. It is clear that the broadening of the central peak is little affected by second order nuclear quadrupole broadening in contrast to the HDHTSC. This is evident in figure 1 where the $^{63}$Cu NMR spectra taken at 8.45 T (filled circles) and 5.6 T (open circles), scaled by the applied field, are also shown. It can be seen that that the widths are nearly the same, indicating that second order nuclear quadrupole effects are small. The weaker broad signal is likely to be due to the +1/2↔+3/2 and -1/2↔-3/2 satellite transitions and a distribution of nuclear quadrupole frequencies. We estimate, from the NMR spectra at different magnetic fields, that the average nuclear quadrupole frequency is less than 3 MHz. This is significantly smaller than that observed in the HDHTSC (16 MHz to 40 MHz [36]). However it is comparable to the upper estimate of the nuclear quadrupole frequency in superconducting samples of the EDHTSC, $Nd_{2-x}Ce_xCuO_4$ (< 2 MHz [15]), and it is consistent with the estimates from two previous studies on $Sr_{1-x}La_xCuO_2$ with x>=0.07 (<~3 MHz [16,18]).

We show in figure 3 that the $^{63}$Cu NMR spectrum from a c-axis aligned sample at 9 T with $c \perp B$ shifts to lower frequency for temperatures below $T_c$ and there is a significant increase in the linewidth as can be seen in the insert to figure 3. An increase in the $^{63}$Cu NMR linewidth is also observed in the HDHTSC and it has been associated with vortices in a type II superconductor and the formation of a flux lattice at low temperatures [37]. The corresponding $^{63}$Cu NMR shift, $\delta_{ab}$, is plotted in figure 4, where it can be seen that $\delta_{ab}$ is temperature independent in the normal state as reported by Mikhalev *et al.* [18] from measurements on an unaligned powder sample. Imai *et al.* [16] also noted that the $\delta_{ab}$ was temperature independent but the shift data were not presented. The $^{63}$Cu NMR shift can be understood by noting that, for axial symmetry, the total NMR shift, $\delta_\alpha$, of the central transition in an external field *B* directed in the α-direction with respect to the crystal frame shift can be written as,

$$\delta_\alpha = \frac{\Delta \nu_\alpha}{\nu_0} = {}^{63}K_{s,\alpha}(T) + {}^{63}K_{orb,\alpha} + K_{dia,\alpha} + \frac{\Delta \nu_{Q,\alpha}^{(2)}}{\nu_0} \qquad (4)$$

where, $\alpha$ is ab for $c\|B$ or c for $c\perp B$, $\Delta\nu_\alpha$ is the shift in frequency from the Larmor frequency, $\nu_0$, $^{63}K_{s,\alpha}(T)$ is the temperature dependent Knight shift, $^{63}K_{orb,\alpha}$ is the temperature independent orbital shift, $K_{dia,\alpha}$ is the diamagnetic shift due to the superconducting screening currents and $\Delta\nu_{Q,\alpha}^{(2)}/\nu_0$ is the second order nuclear quadrupole shift. The diamagnetic shift decreases with increasing magnetic field due to increased flux penetration. We estimate this shift in a manner similar to Zheng *et al.* [38], where the previous estimate was made for the HDHTSC, YBa$_2$Cu$_4$O$_8$, and found to be close to the experimental value. Using the experimentally determined values of the penetration depth and the superconducting coherence length [21], we find that $K_{dia,\perp} \approx -0.007\%$ at 9 T and for low temperatures where the flux solid occurs.

The second order quadrupole term in equation (4) is large in the HDHTSC because of the large electric field gradient at the Cu nucleus and it can be the dominant term, leading to a NMR shift of up to 3.5 % for $c\perp B$. The subtraction of the large second order quadrupole, and the small diamagnetic term, at low temperatures leads to the orbital shift which, for the HDHTSC, is ~0.22 % in YBa$_2$Cu$_3$O$_7$, ~0.22 % in Tl$_2$Ba$_2$CuO$_{6+\delta}$, and ~0.30 % in La$_{2-x}$Sr$_x$CuO$_4$ [36]. These values are very similar to the low temperature NMR shift from our Sr$_{0.9}$La$_{0.1}$CuO$_2$ sample which, after correction from the small diamagnetic shift, is 0.21 % for $c\perp B$.

A temperature independent $^{63}K_{s,\perp}(T)$ for T>50 K is not generally observed in the HDHTSC, except in the slightly overdoped region. For underdoped HDHTSC, it is found that $^{63}K_s(T)$ decreases with decreasing temperature in the normal state, which has been attributed to a normal-state pseudogap [39,40]. In the heavily overdoped region, $^{63}K_s(T)$ increases with decreasing temperature in the normal state, which may be due to the proximity to a van Hove singularity. A varying density of states (DOS) within ~$2k_BT$ of the Fermi level, $E_f$, affects the Knight shift because the Knight shift is proportional to the static spin susceptibility, which can be expressed as, $\chi_s(T) \propto \int_{-\infty}^{\infty} N_s(\varepsilon)(-\partial f(\varepsilon)/\partial\varepsilon)d\varepsilon$ where $\varepsilon = E - E_f$, $N_s(\varepsilon)$ is the DOS and $f(\varepsilon)$ is the Fermi function. Consequently, the absence of a decrease in $^{63}K_{s,\perp}(T)$ with decreasing temperature indicates that there is no significant normal state pseudogap in the EDHTSC, Sr$_{0.9}$La$_{0.1}$CuO$_2$. Rather, the Fermi surface is flat at least for energies within ~$2k_BT$ of the Fermi energy.

The Knight shift above 50 K is estimated from figure 4, and by accounting for the small diamagnetic correction, to be 0.09 %. This is significantly less than that observed in optimally doped HDHTSC. For example, $^{63}K_s(300K)$ is ~0.36 % for YBa$_2$Cu$_3$O$_7$ (T$_c$~90 K) and ~0.40 % for La$_{1.85}$Sr$_{0.15}$CuO$_4$ (T$_c$=38 K) [36]. Using the Knight shift and the known HDHTSC Cu hyperfine coupling constants, we estimate that the static spin susceptibility in the normal state for Sr$_{0.9}$La$_{0.1}$CuO$_2$ is $1.2 \times 10^{-5}$. This is slightly less than the measured susceptibility in our sample of $1.6 \times 10^{-5}$, indicating that the smaller Knight shift can not be accounted for by a large reduction in the hyperfine coupling constants. Consequently, the smaller Knight shift implies that the DOS at the Fermi level, $N(E_f)$, in Sr$_{0.9}$La$_{0.1}$CuO$_2$ is less than ~1/4 of $N(E_f)$ in YBa$_2$Cu$_3$O$_7$ and La$_{1.85}$Sr$_{0.15}$CuO$_4$. This, along with the flat Fermi surface within $\sim 2k_BT$ of the Fermi energy, implies that models developed to account for superconductivity that are based on differences in the band structure (e.g. von Hove singularity, 2 band models etc.) do not appear to be relevant to the HTSC. The significantly smaller $N(E_f)$ found in Sr$_{0.9}$La$_{0.1}$CuO$_2$ when compared with the HDHTSC is likely to be due to the different band structures.

It can be seen in the insert to figure 4 that the Knight shift decreases for temperatures below T$_c$ due to the decreased DOS near $E_f$ in the superconducting state. However, the Knight shift can not be fitted by assuming either a s-wave superconducting order parameter (dotted curve) or a d-wave superconducting order parameter (dashed curve) using the experimentally observed $2\Delta/k_BT_c = 7$ [22]. The effect of the small temperature dependent diamagnetic correction is not large enough to account for this discrepancy. A similar problem exists with the HDHTSC where the Knight shift can be nearly temperature independent for low temperatures [36] although it should be linearly dependent on temperature due to the d-wave symmetry of the superconducting order parameter. It has been suggested that the nearly temperature independent Knight shift observed in some of the HDHTSC for T<<T$_c$ is due to a finite DOS at the Fermi level [36,41]. We show by the solid curve in the insert to figure 4, the theoretical $\delta_{ab}$ obtained by assuming a d-wave superconducting order parameter and a similar residual DOS with $N_{res}(E)/N_0 = 0.25$, where $N_0$ is the DOS in the normal state. It can be seen that that does give a better representation of the data.

The $^{63}$Cu spin lattice relaxation after partial inversion is shown in the insert of figure 5. Here we plot the data as $(M_0 - M(\tau))/M_0$ at 23 K. The recovery after partial inversion can be fitted to equation (1) using $1 < \Gamma < 2$ with the coefficients $A = 0.4$, $B = 0$ and $C = 0.6$. This can be contrasted with the report by Mikhalev *et al.* [18] where the coefficients used were $A = 0.2$, $B = 0$ and $C = 0.8$ but the coefficient in one of the exponential exponents was not reported. The values used by Imai *et al.* [16] were $A = 0.1$, $B = 0$ and $C = 0.9$. The resultant $1/^{63}T_1T$ is plotted in figure 5 where it can be seen that $1/^{63}T_1T$ increases with decreasing temperature in the normal-state even though the NMR shift is independent of temperature in the normal state, which means that the Korringa relation, $^{63}T_1T\,^{63}K_s$ =constant, is not observed in this EDHTSC. A similar departure from Korringa behavior is observed in the HDHTSC. We note that our $1/^{63}T_1T$ are larger than those previously observed. For example $1/^{63}T_1T$ at 300 K is reported to be ~ 7 s$^{-1}$K$^{-1}$ by Imai et al. and ~3 s$^{-1}$K$^{-1}$ by Mikhalev et al. compared with ~11 s$^{-1}$K$^{-1}$ in this study.

It is apparent in figure 5 that $1/^{63}T_1T$ in the normal-state is remarkable similar to that observed in the optimally and overdoped HDHTSC, $La_{2-x}Sr_xCuO_4$ (x>=0.15). The expected behavior of $1/^{63}T_1T$ in the normal state can be understood by noting that for metallic and magnetic systems [42], $(T_1T)^{-1} = (1/2)\eta k_B \gamma_n^2 \sum_q |A(q)|^2 \chi''(q,\omega_0)/\eta\omega_0$, where $|A(q)|$ is the form factor containing onsite and transferred hyperfine coupling constants, $\gamma_n$ is the nuclear gyromagnetic ratio and $\chi''(q,\omega_0)$ is the imaginary part of the dynamical spin susceptibility at the Larmor frequency, $\omega_0/2\pi$. In the case of the HDHTSC, there is clear evidence from inelastic neutron scattering studies that antiferromagnetic fluctuations exist. Thus, it is assumed that $1/^{63}T_1T$ in the HDHTSC is dominated by coupling to the antiferromagnetic spin fluctuation spectrum, leading to an increase in $1/^{63}T_1T$ with decreasing temperature [4]. Hence, the observation that the temperature dependence and absolute values of $1/^{63}T_1T$ from $Sr_{0.9}La_{0.1}CuO_2$ are comparable to the HDHTSC is consistent with both the HDHTSC and the EDHTSC having a similar spin fluctuation spectrum.

We note that for most underdoped HDHTSC, $1/^{63}T_1T$ is found to start to decrease for temperatures below a characteristic temperature T* that is above $T_c$, which has been attributed to a normal-state pseudogap [43,44]. The absence of a similar decrease in $1/^{63}T_1T$ plotted in figure 3 is consistent with there being no significant normal-state pseudogap near $q=(\pi,\pi)$ in the EDHTSC, $Sr_{0.9}La_{0.1}CuO_2$.

Below $T_c$ we find that $1/^{63}T_1$, plotted in figure 6, can not be interpreted in terms of an isotropic s-wave superconducting order parameter. For an isotropic s-wave superconducting order parameter, coherence effects should cause $1/^{63}T_1$ to initially increase for temperatures just below $T_c$ and then rapidly decrease where $1/^{63}T_1 \propto \exp(-\Delta/k_BT)$ is expected for T<<$T_c$. It can be seen in figure 6 that, in agreement with the report by Imai et al., there is no evidence of a coherence peak in $1/^{63}T_1$ near $T_c$ for B=9 T. Furthermore, it is apparent that $1/^{63}T_1 \propto T$ when T<<$T_c$. The values of $1/^{63}T_1$ at low temperature are significantly greater than those expected from a superconductor with a d-wave superconducting order parameter and in zero field. The predicted temperature dependence of $1/^{63}T_1$ in zero field and for a d-wave superconductor can be obtained from $1/^{63}T_1 \propto T\int N(E)f(E)[1-f(E)]dE$. The resultant curve is plotted in figure 6 (dashed curve) where it can be seen that $1/^{63}T_1$ should be proportional to $T^3$ for T<<$T_c$. This is observed in the HDHTSC, YBa$_2$Cu$_3$O$_{7-\delta}$, from zero field NMR measurements but not in the La$_{2-x}$Sr$_x$CuO$_4$ where $1/^{63}T_1$ is proportional to T for optimally and overdoped samples [36]. A linear dependence of $1/^{63}T_1$ for T<<$T_c$ can arise from a residual DOS as shown by the dotted curve in figure 6. Here we use the same DOS used to analyze the NMR shift plotted in figure 4 (dotted curve), where $N_{res}(E)/N_0 = 0.25$. Increasing the residual DOS can lead to a better representation of the data as shown in figure 6 where we use $N_{res}(E)/N_0 = 0.5$ (solid curve). However, this implies a significantly larger residual DOS than found in some of the pure HDHTSC. It is possible that there is an additional relaxation contribution from the fluxoids, which can lead to a low temperature spin lattice relaxation rate that is proportional to T [45].

**Conclusion**

In conclusion, we find that the temperature dependence of the normal-state thermopower from the infinite layer EDHTSC, Sr$_{0.9}$La$_{0.1}$CuO$_2$, is different from that observed in the other HTSC. In particular, dS(T)/dT is positive for temperatures above 120 K, which is not observed in HDHTSC that only contain CuO$_2$ planes. However, S(T) can be fitted to the thermopower model developed for the HDHTSC where the thermopower is modeled in terms of a diffusion term and a phonon drag term. Consequently, a previous interpretation of the thermopower data from Sr$_{0.9}$La$_{0.1}$CuO$_2$ thin films in terms of phonon mediated pairing is inappropriate. The $^{63}$Cu Knight shift from the EDHTSC, Sr$_{0.9}$La$_{0.1}$CuO$_2$ is much smaller than from the HDHTSC with the same $T_c$, which we interpret in terms of a much smaller DOS at the Fermi level. Furthermore, the $^{63}$Cu Knight shift is temperature independent in the normal state indicating a flat DOS at least for energies $\sim 2k_BT$ from the Fermi level. These imply that theories developed to explain the origin of superconductivity in the HTSC that are based on the simple band structure effects can not be applied to this EDHTSC. We find that the temperature dependence and absolute values of $1/^{63}T_1T$ in the normal state are similar to those of the optimally doped and overdoped HDHTSC, La$_{2-x}$Sr$_x$CuO$_2$, implying that the electron and hole doped HTSC have the same dynamical spin susceptibility. Below $T_c$ the $^{63}$Cu NMR shift and $1/^{63}T_1$ decrease more slowly than expected for an isotropic s-wave superconductor. We show that this can be interpreted in terms of a d-wave superconducting order parameter, a residual DOS and additional relaxation by fluxoids. Thus, the significantly smaller DOS, the flat DOS within $2k_BT$ of the Fermi level, but comparable dynamical spin susceptibility and symmetry of the superconducting order parameter would appear to suggest that the pairing mechanism is mediated by antiferromagnetic spin fluctuations as suggested by some of the models of the HDHTSC.

**Acknowledgements**

We acknowledge useful discussion with M. Mehring and J. Haase and funding support from the New Zealand Marsden Fund (GVMW, HJT), the Alexander von Humboldt Foundation (GVMW), the UK EPSRC (RD and AH) and the Ministry of Science and Technology of Korea through the Creative Research Initiative Program (C. U. Jung, Min-Seok Park and Sung-Ik Lee).

**References**

[1] Z. X. Shen, W.E. Spicer, D. M. King, D.S. Dessau, B. O. Wells, Science **267**, 343 (1995).

[2] J. R. Kirtley, C.C. Tsuei, J. Z. Sun, C. C. Chi, L. S. Yujahnes, A. Gupta, M. Rupp, M. B. Ketchen, Nature (London) **373**, 225 (1995).

[3] A. G. Loeser, Z. X. Shen, D. S. Dessau, D. S. Marshall, C. H. Park, P. Fournier, A. Kapitulnik, Science **273**, 325 (1996).

[4] A. J. Millis, H. Monien and D. Pines, Phys. Rev. B **42**, 167 (1990).

[5] S. Kashawaya, T. Ito, K. Oka, S. Ueno, H. Takashima, M. Koyanagi Y. Tanaka and K. Kajimura Phys. Rev. B **57**, 8680 (1998).

[6] L. Alff, S. Meyer, S. Kleefisch, U. Schoop, A. Marx, H. Sato, M. Naito, and R. Gross, Phys. Rev. Lett. **83**, 2644 (1999).

[7] J. Skinta, T. R. Lemberger, T. Greibe and M. Naito, cond-mat/0108545 and cond-mat/0107122.

[8] C. C. Tsuei and J. R. Kirtley, Phys. Rev. Lett. **85**, 182 (2000).

[9] R. Prozorov, W. Giannetta, P. Fournier and R. L. Greene, Phys. Rev. Lett. **85**, 3700 (2000).

[10] N. P. Armitage, D. H. Lu, D. L. Feng, C. Kim, A. Damascelli, K. M. Shen, F. Ronning, Z.-X. Shen, Y. Onose, Y. Taguchi, and Y. Tokura , Phys. Rev. Lett. **86**, 1126 (2001).

[11] S. Kleefisch B. Welter, A. Marx, L. Alff, and R. Gross M. Naito, Phys. Rev. B **63**, 100507 (2001).

[12] A. Biswas, P. Fournier, V. N. Smolyaninova, R. C. Budhani, J. S. Higgins, and R. L. Greene, Phys. Rev. B **64**, 104519 (2001).

[13] N. P. Armitage, D.H. Lu, C. Kim, A. Damascelli, K.M. Shen, F. Ronning, D.L. Feng, P. Bogdanov, Z.-X. Shen, Y. Onose, Y. Taguchi, Y. Tokura, P.K. Mang, N. Kaneko, M. Greven, Phys. Rev. Lett. **87**,  147003 (2001).

[14] E. J. Singley, D. N. Basov, K. Kurahashi, T. Uefuji and K. Yamada, Phys. Rev. B **64**, 224503 (2001).

[15] S. Kambe, H. Yasuoka, H. Takagi, S. Uchida and Y. Tokura, J. Phys. Soc. Japan **60**, 400 (1991).

[16] T. Imai, C. P. Slichter, J. L. Cobb and J. T. Markert, J. Phys. Chem. Solids **56**, 1921 (1995).

[17] T. Imai, C. P. Slichter, K. Yoshimura, M. Katoh, K. Kosuge, J. L. Cobb, and J. T. Markert, Physica C **235-240**, 1627 (1994).

[18] K. Mikalev, K. Kumagai, Y. Furukawa, V. Bobrovski, T. D'yachkova, N. Kad' irova and A. Gerashenko, Physica C **304**, 165 (1998).

[19] M. G. Smith, A. Manthiram, J. Zhou, J. B. Goodenough and J. T. Markert, Nature **351**, 549 (1991).

[20] J. H. Schön, M. Dorget, F. C. Beuran, X. Z. Zu, E. Arushanov, C. Deville Carellin and M. Laguës, Nature **414**, 434 (2001).

[21] Mun-Seog Kim, C. U. Jung, J. Y. Kim, Jae-Hyuk Choi, Sung-Ik Lee, cond-mat/0102420.

[22] C. –T. Chen, P. Seneor, N. –C. Yeh, R. P. Vasquez, C. U. Jung, J. Y. Kim, Min-Seok Park, Hung-Jeong Kim and S. I. Lee, (submitted to Science).

[23] C. U. Jung, J. Y. Kim, S. M. Lee, M. S. Kim, Y. Yao, S. Y. Lee, S. I. Lee, D. H. Ha, Physica C 364, 225 (2001).

[24] C. U. Jung, Mun-Seog Kim, C. U. Jung, J. Y. Kim, Jae-Hyuk Choi, Sung-Ik Lee, cond-mat/0106441.

[25] E. C. Jones, D. P. Norton, B. C. Sales, D. H. Lowndes and R. Feenstra, Phys. Rev. B **52**, 743 (1995).

[26] C. U. Jung, Min-Seok Park, W. N. Kang, Mun-Seog Kim, Kijoon H. P. Kim, S. Y. Lee, and Sung-Ik Lee, Appl. Phys. Lett. **78**, 4157 (2001).

[27] E. R. Andrew and D. P. Tunstall, Proc. R. Soc. London **78**, 1 (1961).

[28] A. B. Kaiser and C. Uher in *High Temperature Superconductors* edited by A. V. Narlokar (Nova, New York, 1991), vol. **7**.

[29] C. R Varoy, H. J. Trodahl, R. G. Buckley and A. B. Kaiser, Phys. Rev. B **46**, 463 (1992).

[30] S. D. Obertelli, J. R. Cooper and J. L. Tallon, Phys. Rev. B **46**, 14 928 (1992).

[31] C. Bernhard and J. L. Tallon, Phys. Rev. B **54**, 10201 (1996).

[32] H. J. Trodahl, Phys. Rev. B **51**, 6175 (1995).

[33] R. Carter and P. A. Schroeder, J. Phys. Chem. Solids **31**, 2374 (1970).

[34] F. J. Blatt, P. A. Schroeder, C. Foiles and D. Greig, *Thermoelectric Power in Metals* (Plenum, New York, 1976).

[35] R. D. Barnard, *Thermoelectricity in Metals and Alloys* (Taylor and Francis, London, 1972).

[36] For a review see, K. Asayama, Y. Kitaoka. G. Q. Zheng and K. Ishida, *Progress in Nuclear Magnetic Resonance Spectroscopy* **28**, 221 (1996).

[37] A. P. Reyes, X. P. Tang, H. N. Bachman, W. P. Halperin, J. A. Martindale and P. C. Hammel, Phys. Rev. B **55** 14737 (1997).

[38] G. Q. Zheng W. G. Clark Y. Kitaoka and K. Asayama Y. Kodama P. Kuhns and W. G. Moulton , Phys. Rev. B **60**, 9947 (1999).

[39] M. Mehring, Appl. Magn. Reson. **3**, 383 (1992).

[40] G. V. M. Williams, J. L. Tallon, E. M. Haines, R. Michalak, and R. Dupree, Phys. Rev. Lett. **78**, 721 (1997).

[41] S. Ohsugi, Y. Kitaoka, K. Ishida, G. Zheng and K. Asayama, J. Phys. Soc. Japan **63**, 700 (1994).

[42] T. Moriya, J. Phys. Soc. Japan **18**, 516 (1963).

[43] M. Bankay, M. Mali, J. Roos, and D. Brinkmann, Phys. Rev. B **50**, 6416 (1994).

[44] G. V. M. Williams, J. L. Tallon and J. W. Loram, Phys. Rev. B **58**, 15053 (1998).

[45] N. J. Curro, C. Milling, J. Haase and C. P. Slichter, Phys. Rev. B **62**, 3473 (2000).

**Figures**

**Figure 1:** Plot of the thermopower against temperature. Also shown is the low temperature (dashed curve) and high temperature (solid curve) fit to equation 3 in the text. Insert: Plot of the zero-field-cooled (solid curve) and field-cooled (dashed curve) dc susceptibility for an applied magnetic field of 2 mT.

**Figure 2:** Plot of the $^{63}$Cu NMR spectra at 293 K for an applied magnetic field of 14.1 T (solid up triangles). Also shown is the NMR spectra at 8.45 T (solid circles) and 5.6 T (open circles) scaled by the respective Larmor frequencies.

**Figure 3:** Plot of the $^{63}$Cu NMR spectra at 10 K (dot dashed curve), 20 K (dashed curve), 30 K (doted curve) and 40 K (solid curve) for c⊥B. Insert: Plot of the corresponding linewidths.

**Figure 4:** Plot of the $^{63}$Cu NMR shift $\delta_{ab}$ for c⊥B and 9 T. Insert: Plot of the $^{63}$Cu NMR shift for c⊥B and over a smaller temperature range. Also shown is the expected temperature dependence for an isotropic s-wave superconducting order parameter (dotted curve), a d-wave superconducting order parameter (dashed curve) and a d-wave superconducting order parameter with a residual DOS of $N_{res}(E)/N_0 = 0.25$ (solid curve) where $2\Delta/k_B T_c = 7$ [22].

**Figure 5:** Plot of $1/^{63}T_1 T$ against temperature at 9 T from $Sr_{0.9}La_{0.1}CuO_2$ (solid circles) and $La_{1.76}Sr_{0.24}CuO_4$ (crosses [41]), $La_{1.80}Sr_{0.20}CuO_4$ (circles [41]) and $La_{1.85}Sr_{0.15}CuO_4$ (up triangles [41]). Insert: Plot of $(M_0 - M(\tau))/M_0$ from $Sr_{0.9}La_{0.1}CuO_2$ at 23 K.

**Figure 6:** Plot of $1/^{63}T_1$ against temperature (circles). Also shown is $1/^{63}T_1$ expected for a d-wave superconducting order parameter (dashed curve), a d-wave superconducting order parameter with $N_{res}(E)/N_0 = 0.25$ (dotted curve) and $N_{res}(E)/N_0 = 0.5$ (solid curve).

Figure 1
Phys. Rev. B
Williams *et al.*

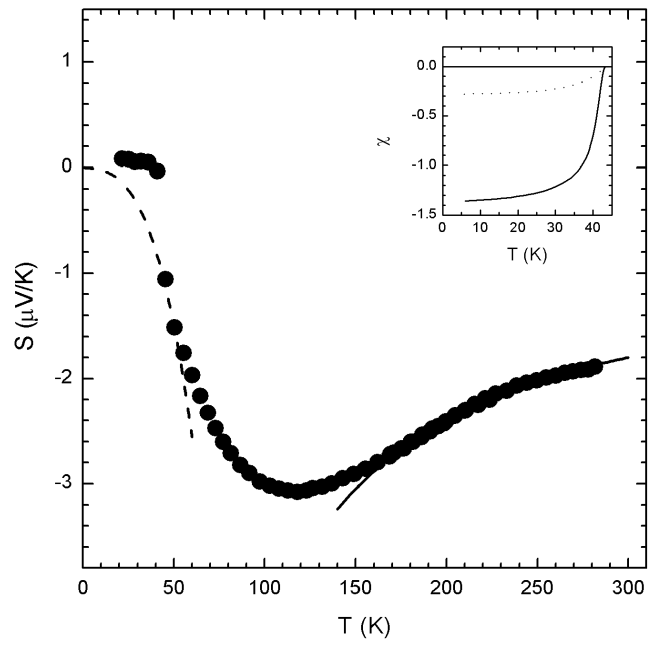

Figure 2
Phys. Rev. B
Williams *et al.*

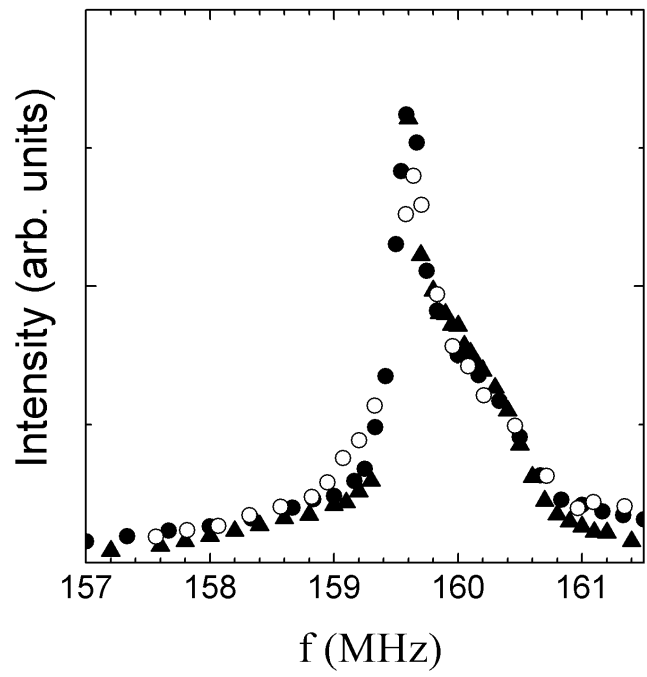

Figure 3
Phys. Rev. B
Williams *et al.*

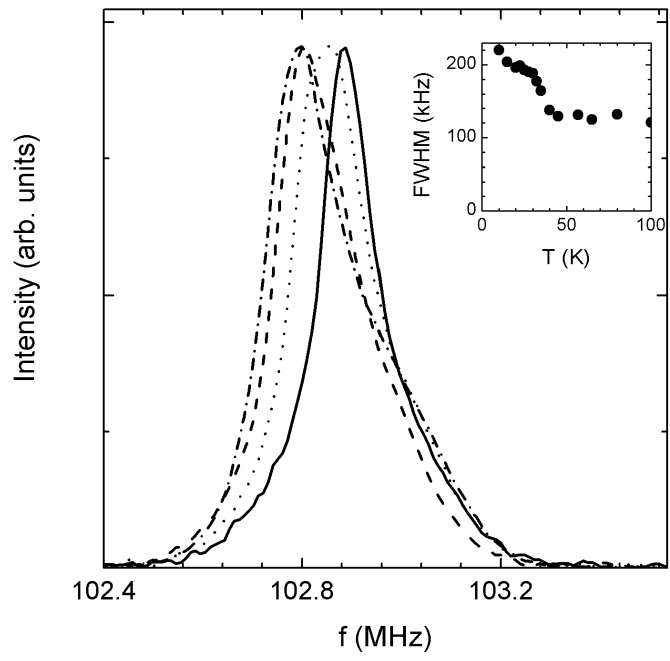

Figure 4
Phys. Rev. B
Williams *et al.*

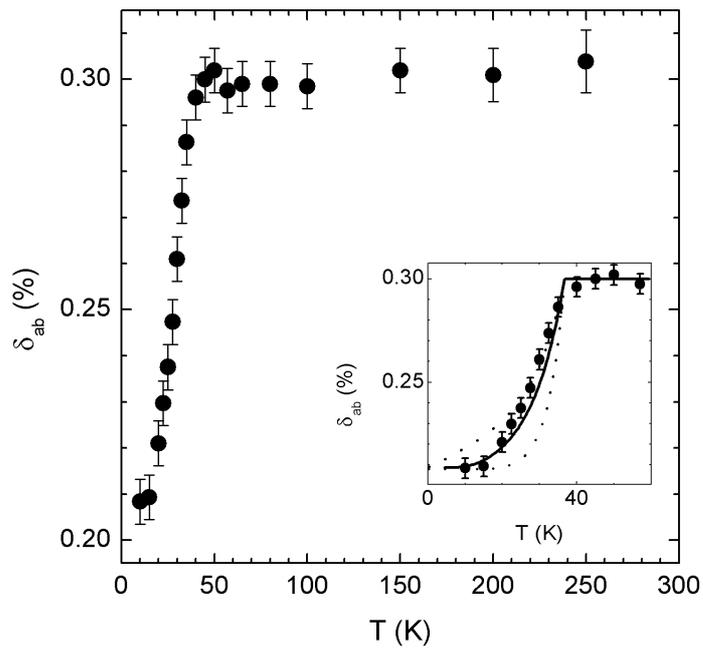

Figure 5
Phys. Rev. B
Williams *et al.*

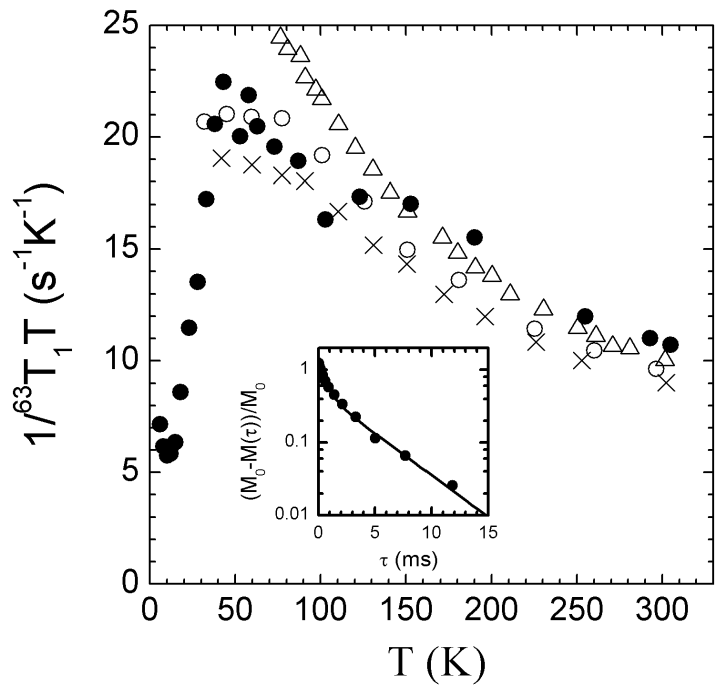

Figure 6
Phys. Rev. B
Williams *et al.*

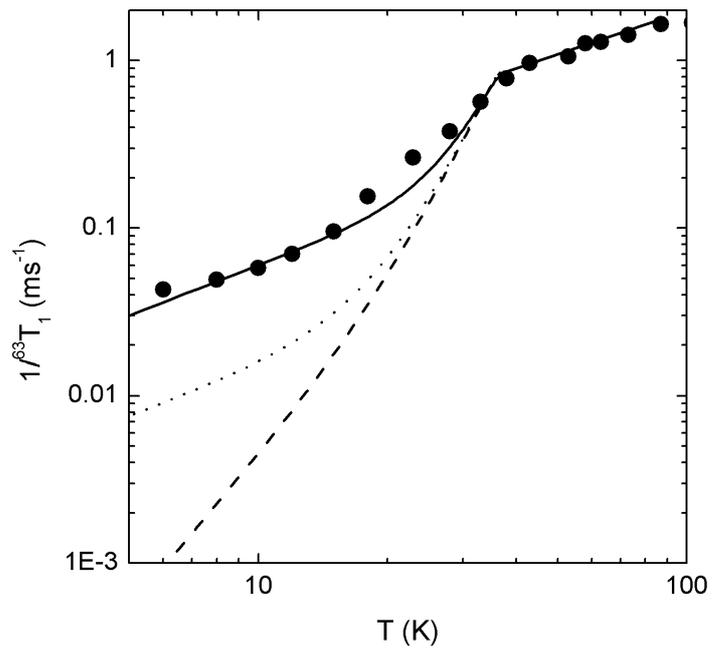